\documentclass[prl,aps,twocolumn,showpacs,floatfix,groupedaddress]{revtex4-1}

\usepackage{graphicx}
\usepackage{dcolumn}
\usepackage{bm}
\usepackage{latexsym}
\usepackage{amsmath}
\usepackage{amssymb}
\usepackage[latin1]{inputenc}
\usepackage[usenames]{color}
\usepackage{wasysym}
\usepackage[usenames,dvipsnames]{xcolor}
\usepackage{multirow}
\usepackage{stackengine}

\begin{document}

\title{The simultaneous discharge of liquid and grains from a silo}

\author{A. M. Cervantes-\'Alvarez, S. Hidalgo-Caballero and F. Pacheco-V\'azquez}

\affiliation{Instituto de F\'isica, Benem\'erita Universidad Aut\'onoma de Puebla, Apartado Postal J-48, Puebla 72570, Mexico}

\date{\today}

\begin{abstract}
The flow rate of water through an orifice at the bottom of a container depends on the hydrostatic pressure whereas for a dry granular material is nearly constant. But what happens during the simultaneous discharge of grains and liquid from a silo? By measuring the flow rate as a function of time, we found that: (i) different regimes appear, going from constant flow rate dominated by the effective fluid viscosity to a hydrostatic-like discharge depending on the aperture and grain size, (ii) the mixed material is always discharged faster than dry grains but slower than liquid, (iii) for the mixture, the liquid level drops faster than the grains level but they are always linearly proportional to one another, and (iv) a sudden growth in the flow rate happens during the transition from a biphasic to a single phase discharge. These results are associated to the competition between the decrease of hydrostatic pressure above the granular bed and the hydrodynamic resistance. A model combining the Kozeny-Carman, Bernoulli and mass conservation equations is proposed and the numerical results are in good agreement with experiments.  
\end{abstract}


\maketitle

The clepsydra was an ancient device used to measure the passage of time based on the slow discharge of water through an orifice at the bottom of a graduated vessel. Its origin is unknown but presumably appeared in China about six thousands years ago \cite{Needham2000}. The markings in this antiquity are non-uniformly separated since the flow rate depends on the hydrostatic pressure due to the water column above the orifice. On the other hand, the clepsammia, or sandglass,  appeared millennia later probably in Alexandria \cite{Benthuysen1870} although evidence of its origin only dates from the middle ages \cite{Balmer1978}.  
In this device, the flow rate of grains through the bottleneck is independent of the granular column height because the stress acting on the material is redirected towards the container walls through contact force chains. This fact attracted scientists during the last 60 years, being the discharge of dry granular materials from hoppers widely investigated \cite{Beverloo1961,Mills1996,Samadani1999,Zuriguel2005,Mankoc2007,Unac2012,Dorbolo2013,Pacheco2017}.

The flow rate $Q$ of dry grains through an orifice can be estimated using the Beverloo equation \cite{Beverloo1961}:
\begin{equation}
Q=C \rho \sqrt{g}\left(D-kd \right)^{5/2}, 
\label{eq1a}
\end{equation} 
where $D$ and $d$ are the aperture and particle diameters, $C$ and $k$ are dimensionless fitting parameters related to friction and particle shape, $\rho$ is the material density and $g$ is the acceleration of gravity. The above equation must be modified by introducing an exponential factor when the ratio $D/d$ is considerably increased \cite{Mankoc2007}. More recently, experiments performed at different $g$'s proved that the square root scaling proposed by Beverloo is relevant \cite{Dorbolo2013}.
It has also been found that the initial packing fraction of the granular column does not affect considerably the flow rate \cite{Huang2006,Ahn2008,Pacheco2017} because the material under discharge is fluidized before reaching the silo aperture\cite{Pacheco2017}. Even in unconventional systems of repelling particles, $Q$ remains constant during the discharge \cite{Lumay2015,Hernandez2017}. 

In all the above studies, the interstitial medium is air and its presence was neglected or assumed to be equivalent to the discharge of dry grains in vacuum. Only recently, the discharge of silos totally submerged in water was considered \cite{Wilson2014,Koivisto2017,Koivisto2017-2}. Under these conditions, the flow rate is not constant unless the filling height is very large, and increases as the hopper empties. An unexpected surge in the flow rate appears near the end of the discharge, which is attributed to a pumping effect produced by the interstitial fluid moving faster than the grains. In flow-controlled experiments the surge disappears and the flow rate becomes constant \cite{Koivisto2017}.


In this article, we study the simultaneous discharge of glass beads and water from a cylindrical silo. An important difference with previous works is that our system is not underwater but it hangs freely in air from a force sensor that allows us to determine the flow rate throughout the emptying process. We found that $Q$ can be constant, increase or  decrease as the silo empties depending on $d$, $D$ and on the size of the liquid column above the granular bed. The surge previously reported in underwater systems is also observed and the difference between the flow rate of the mixed phase and the liquid phase reveals a change in the effective viscosity of the fluid. The liquid level always decreases proportionally to the grains level, and this dependence is introduced in the Kozeny-Carman and Bernoulli equations to model the discharge. 

\begin{figure*}[ht!]
\begin{center}
\topinset{a) \hspace{3.7cm} b) \hspace{5cm} c)}{\includegraphics[width=15cm]{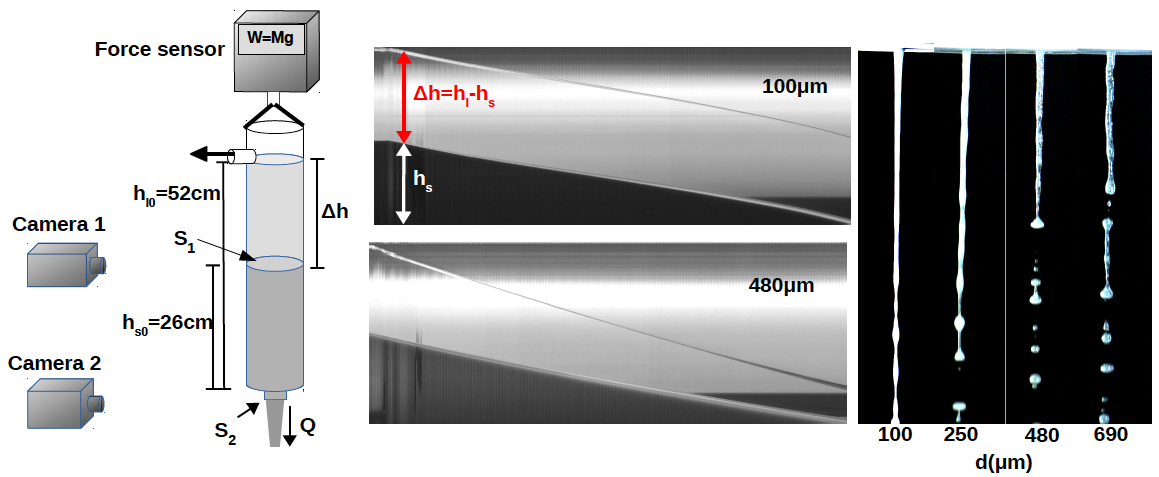}}{0.0cm}{-1.5cm}
\end{center}
\vspace*{-0.4cm}
\caption{\small (Color online): a) Experimental setup. b) Montages of 10-pixel width snapshots obtained from videos filmed with camera 1 at 25 fps for $100 \mu$m (top) and $480 \mu$m (bottom). The decrease of water and grains levels can be obtained from these images.  c) Snapshots of the mixture  G$\&$W flowing from apertures of different sizes taken with camera 2. The transition from dripping to jetting as $d$ decreases indicates a considerable increase of the effective viscosity of the mixture \cite{Ambravaneswaran2004}.
}
\label{fig1}
\end{figure*}

\textit{Experimental setup}: The silo consists of  a transparent cylindrical container of 4.4 cm inner diameter and 60 cm high with an interchangeable acrylic flat bottom that allows us to vary the aperture size in four different values $D=3.1, 4.3, 5.4$ and $6.3$ ($\pm 0.01$) mm. The container hangs vertically from a fixed Force Gauge Omega DFG-335 as it is schematically shown in Fig. \ref{fig1}a. This array allows us to measure the system weight as a function of time during the discharge of dry silica grains (density $\rho_g = 2.66\pm0.01$ g/cm$^3$), pure water (density $\rho_l \approx 1$ g/cm$^3$), and grains sedimented in water (G\&W).  The experiments were performed using deionized water at room temperature and four different grain sizes $d$: $100\pm25$, $250\pm40$, $480\pm35$ and $690\pm70$ $\mu$m for each value of $D$.

Before each experiment, the orifice is blocked with duct tape, the silo is hanged from the force gauge and filled with grains of a desired size. In the case of G\&W, the container is first filled with water up to a height $h_{l0}= 52\pm0.2$ cm and then the grains are poured from the top until reaching a granular column of height $h_{s0}=26\pm0.2$ cm. The displaced water escapes through a hole that was made on the lateral wall at 52 cm from the bottom of the silo to ensure that the water level at the beginning of each experiment is always the same. Once the granular material has settled, the duct tape is removed to start the discharge and the weight is recorded with the force gauge at 5 Hz until the process ends. The experiment was repeated five times for different grain size and apertures, and the equivalent procedure was followed using dry grains and water to compare the flow rate for the three situations. 

The simultaneous discharge was filmed laterally with two cameras, one located in front of the silo to measure the decrease of the granular bed and water levels as a function of time (camera 1), and the other one filming the material flowing at the output (camera 2). Figure \ref{fig1}b shows montages obtained by superposing 10-pixels width vertical lines from snapshots captured with camera 1 for two different grain sizes. It can be noticed that the distance between the water level $h_l$ and the grains level $h_s$ is practically constant for particles of $100$ $\mu$m. However, for larger grains of 480 $\mu$m,  $h_l$ decreases considerably faster than $h_s$. Figure \ref{fig1}c taken with camera 2 shows that the filament of material that comes out starts dripping at shorter distances from the orifice as the grain size increases. The above results for different values of $D$ and $d$ are discussed in the following sections.

\begin{figure}[ht!]
\begin{center}
\topinset{a)}{\includegraphics[width=6.5cm]{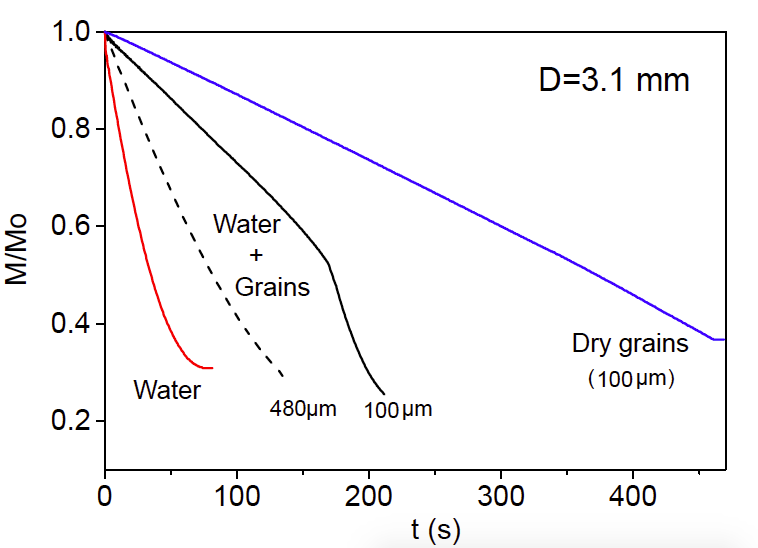}}{0.3cm}{-3.5cm}
\topinset{(b) Grains \hspace*{2.5cm} (c) Water}{\includegraphics[width=9cm]{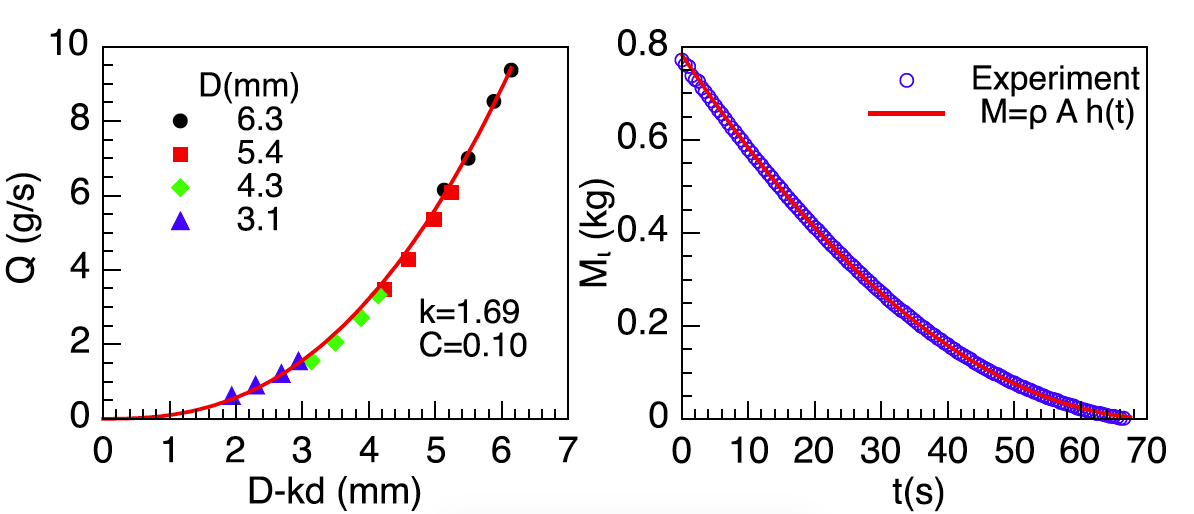}}{0.0cm}{0cm}
\end{center}
\vspace*{-0.5cm}
\caption{\small (Color online): a) Discharge of dry grains (blue line), free water (red line) and grains+water (black lines). b) The flow rate of dry grains (color points) is well described by the Beverloo law given by Eq. \eqref{eq1a} (red line). c) The discharge of water as a function of time (blue points) is well described by Eq. \eqref{eq2} (pink line). }
\label{fig2}
\end{figure}

\begin{figure*}[ht!]
\begin{center}
\topinset{a)}{\includegraphics[width=15cm]{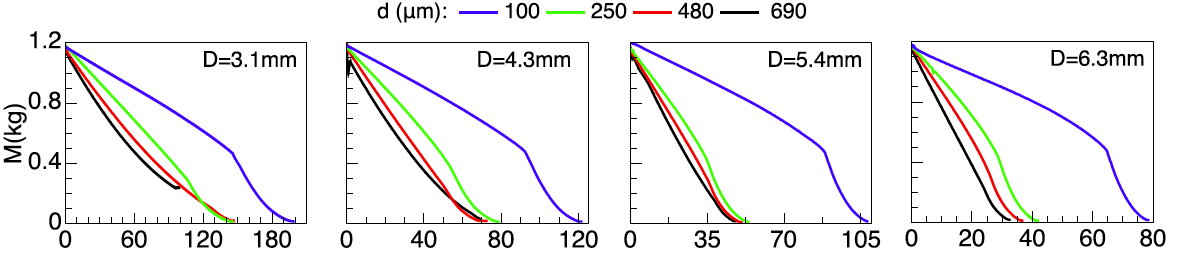}}{0.3cm}{-8.5cm}
\topinset{b)}{\includegraphics[width=15cm]{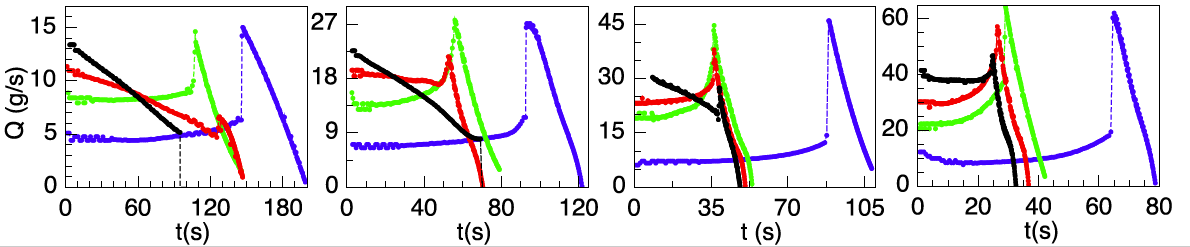}}{0.3cm}{-8.5cm}
\end{center}
\vspace*{-0.5cm}
\caption{\small (Color online): a) Mass $M$ vs time $t$ registered by the force sensor for different values of $D$ and $d$. b) Corresponding flow rate $Q= dM/dt$ as a function of $t$. The break in each curve corresponds to the moment at which the silo runs out of grains and only pure water continues flowing (this transition is indicated with a dashed line in each case).}
\label{fig3}
\end{figure*}

\textit{Results}: In Fig. \ref{fig2}a, the mass registered by the force gauge, $M$, normalized with the initial mass of the full silo, $M_0$, is plotted as a function of time, $t$, for the three situations described above. In the case of dry grains, $M/M_0$ decreases linearly with $t$ (blue line) indicating a constant flow rate of material. For the free water discharge, $M/M_0$ describes a non-linear dependence on time associated with the decrease of hydrostatic pressure (red line). On the other hand, the G\&W discharge displays different dynamics depending on the grains size, going from a practically constant flow regime for grains of 100 $\mu$m (continuous black line) to a more hydrostatic-like regime for 480 $\mu$m (dashed black line). Note that the mixed material is discharged faster than dry grains of the same size but slower than pure water.

\noindent \textit{Individual discharges}: The first two behaviors are expected according to the literature. In fact, the flow rate of dry grains is largely independent of the amount of material above the orifice as it is stated by the Beveloo law. We plot $Q$ vs $(D-kd)$ in Fig. \ref{fig2}b considering all the values of $D$ and $d$ used in our experiments (points) and the data are well fitted by Eq. \eqref{eq1a} (red line). 
On the other hand, to model the discharge of liquid we used the equation of continuity, $S_1 v_1 = S_1 \frac{d h_l}{dt}=S_2 v_2$, combined with the Bernoulli's theorem,
$\rho_l g h_l = \frac{1}{2} \rho_l \left(v_2^2-v_1^2 \right)$, 
where $S_1$ and $v_1$ are the inner cylinder area and the liquid surface velocity, and $S_2=C_d \pi D^2/4$ and $v_2$ are the corresponding effective area and velocity at the output with discharge coefficient $C_d$. By integrating $h_l(t)$ one obtains an expression for the fluid mass $M_l$ remaining in the cylinder, which is given by: 
\begin{equation}
M_l = \rho_l S_1 h_l [t] = \rho_l S_1 \left(\sqrt{h_0}-\frac{1}{2} S_2  \sqrt{\frac{2 g}{S_1^2-S_2^2}} t \right)^2.
\label{eq2}
\end{equation} 
Figure \ref{fig2}c shows the comparison of the numerical solution of Eq. \eqref{eq2} with $C_d\approx 0.9$ (pink line) and the experimental results for the particular case $D=3.1$ mm (blue points), indicating an excellent agreement. 

\begin{figure}[ht!]
\begin{center}
\topinset{a)}{\includegraphics[width=7.0cm]{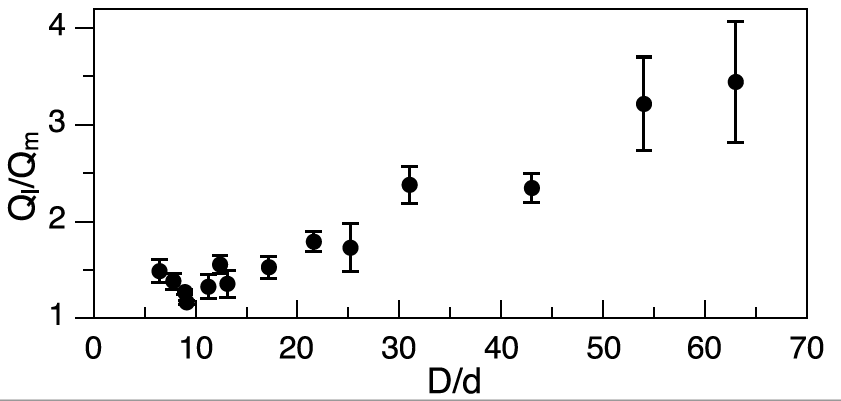}}{0cm}{-3.5cm}
\topinset{b) \hspace{3.2cm} c) }{\includegraphics[width=7.0cm]{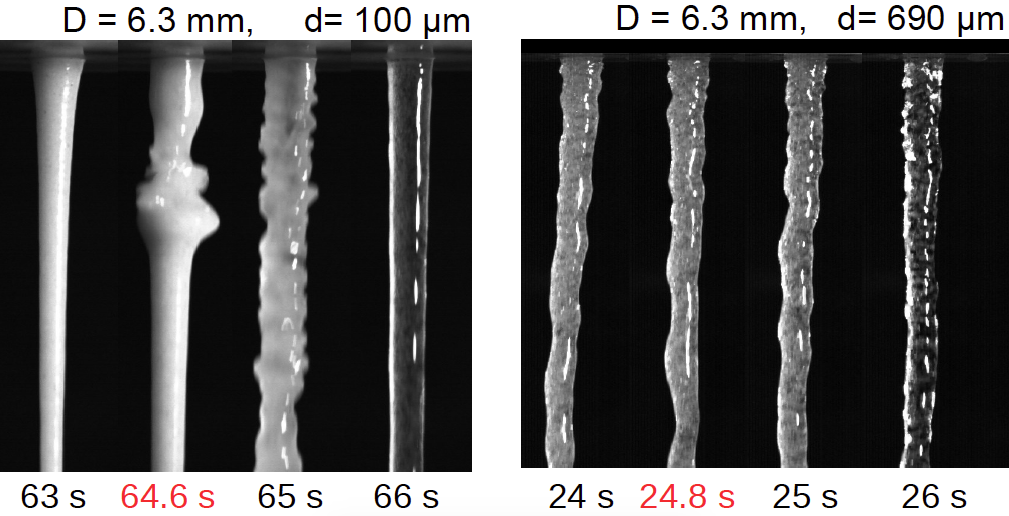}}{0.3cm}{-1.85cm}
\end{center}
\vspace*{-0.6cm}
\caption{\small a) Flow rates quotient for the transition from bi-phasic to single phase discharge as a function of $D/d$. b-c) Grain concentration transition of the jet at the end of the discharge for $d= 100\mu$m (abrupt) and $d=690 \mu$m (gradual). }
\label{fig4}
\end{figure}

\textit{Simultaneous discharge}: Figure \ref{fig3} shows $M$ vs $t$ and the corresponding flow rate $Q$ vs $t$ for all the values of $d$ and $D$ explored in this work. Depending on the combination of these two parameters, $Q$ can decrease, increase or practically remain constant during the emptying process. Moreover, there is a considerable difference between flow rates for the mixed material $Q_m$ and for pure water $Q_l$ when the silo runs out of grains (sudden break in the curves). At that instant, Fig. \ref{fig4}a shows that $Q_l>Q_m$ and the difference increases with $D/d$ being almost four times greater for $D=6.3$ mm and $d=100$ $\mu$m ($D/d\approx63$). For these values, images of the filament taken with camera 2 (Fig. \ref{fig4}b) reveal that a plug of grains at the end of the discharge marks an abrupt reduction of grains concentration in the falling material, and therefore a dramatic decrease in the \textit{effective fluid viscosity} \cite{Ambravaneswaran2004}. The concentration decreases gradually for larger grains and the filament shows more instability (Fig. \ref{fig4}c), which indicates that inertia dominates over viscosity.

\begin{figure}[ht!]
\begin{center}
\includegraphics[width=8.6cm]{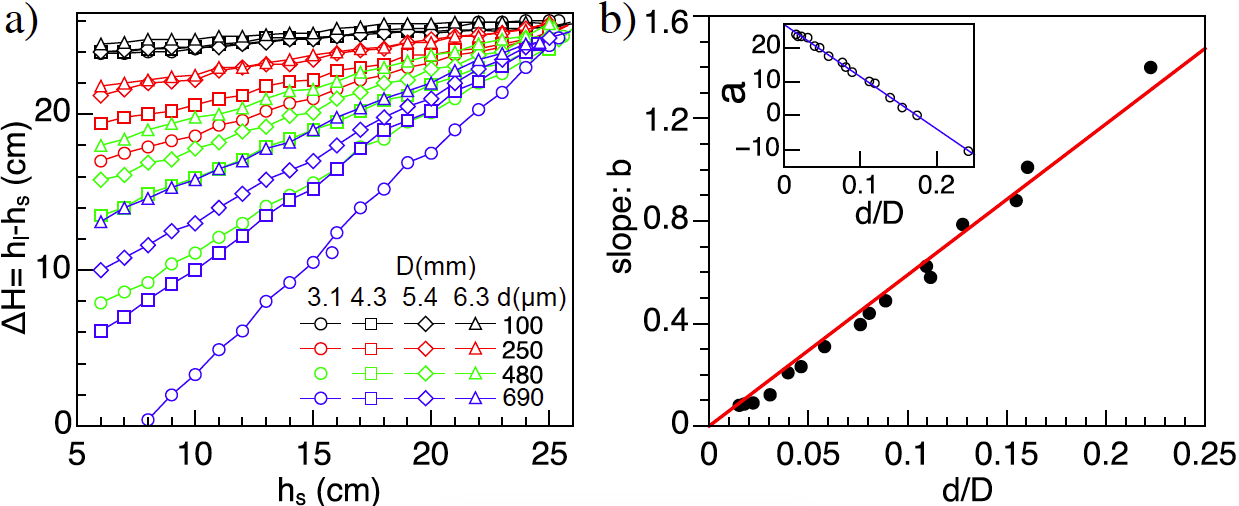}
\end{center}
\vspace*{-0.5cm}
\caption{\small (Color online). a) Difference between water and grains levels $\Delta H$ as a function of the porous substrate height $h_s$ for different combinations of $D$ and $d$. Note that for the smaller grains $\Delta H$ is almost constant during the process. b) Slopes $b$ (main plot) and intercepts $a$ (inset) of the linear behaviors $\Delta H=a+bh_s$ shown in (a) as a function of $d/D$.}
\label{fig5}
\end{figure}

Another remarkable feature of the simultaneous discharge is that the water level measured from the aperture $h_l$ always decreases linearly with the grains level $h_s$. $\Delta H= h_l-h_s$ is plotted as a function of $h_s$ in Fig. \ref{fig5}a.  The slopes $b$ and intercepts $a$ obtained by applying a linear fit of the form $\Delta H= a + b h_s$ to each set of data was plotted as a function of $d/D$ in Fig. \ref{fig5}b (points). The main plot reveals that $b=(5.9\pm0.2)d/D$, and evaluating $a$ in the initial conditions ($h_l=52$ cm and $h_s=26$ cm at $t=0$ s) one obtains $a=26-153.4d/D$ [cm], which is plotted as a blue line in the inset. Since $a$ and $b$ are constants for a fixed $d/D$ and $h_l-h_s = a + bh_s$, from the first derivative over time one obtains that the velocities of the water level $v_l=dh_l/dt$ and the granular bed level $v_s=dh_s/dt$ are related by: 
\begin{equation}
v_l\approx\left(1+ 6\dfrac{d}{D}\right) v_s,
\label{eq3}
\end{equation}
which indicates that the liquid is discharged faster than the grains. When $d\ll D$, the liquid is discharged passively with the granular material. \\

\begin{figure}[ht!]
\begin{center}
\includegraphics[width=8.5cm]{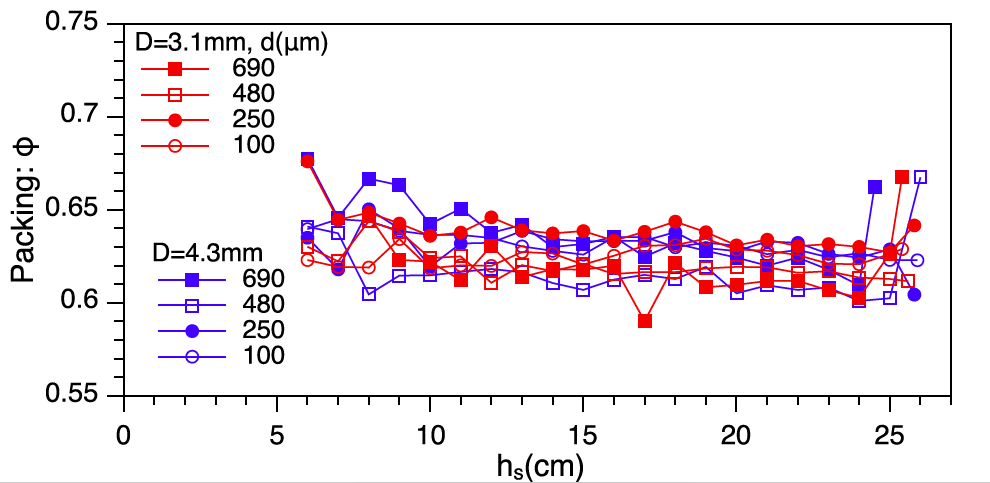}
\end{center}
\caption{\small (Color online): Packing fraction of the immersed grains $\phi$ as a function of $h_s$. The average value is $<\phi> = 0.635\pm 0.018$. Data for $h<5$ cm was not possible to obtain due to set-up limitations but a sudden decrease of $<\phi>$ is expected.  }
\label{fig6}
\end{figure}

\noindent \textit{Modelling the G$\&$W discharge}: let us assume a liquid moving through a porous substrate of thickness $h_s(t)$. This generates a pressure drop $\Delta P$ between the top and the bottom of the granular column that can be estimated using the Kozeny-Carman model:
\begin{equation}
\Delta P = 180 \frac{\eta (1-\epsilon)^2}{d^2 \epsilon^3} h_s v,
\label{eq4}
\end{equation}
where $v$ is the relative velocity between fluid and grains. The porosity $\epsilon$ and the packing fraction of the bed $\phi$ are related by $\phi=1-\epsilon$. Because the porous medium is moving,  $\phi(h)$ is a dynamic parameter that can be varying during the discharge. With the help of the sketch in Fig. \ref{fig1}a, we can notice that the net system mass $M$ is the sum of the mass of grains, the interstitial water and the mass of water above the granular substrate, then:
$M(h_s,\Delta H)= \rho_s S_1 h_s\phi + \rho_l S_1 h_s(1-\phi) + \rho_l S_1 \Delta H$. Solving for $\phi$ one obtains: $$\phi=\frac{M-\rho_l S_1 \Delta H}{(\rho_s-\rho_l)S_1 h_s}.$$ Using the experimental values of $M$, $h_s$ and $\Delta H$ obtained from the force sensor and videos, we plot $\phi(h_s)$ for different apertures and grains size in Fig. \ref{fig6}. Note that the packing is approximately constant during the discharge with an average value $\phi=0.63\pm0.02$.

\begin{figure}[ht!]
\begin{center}
\topinset{a)}{\includegraphics[width=9.1cm]{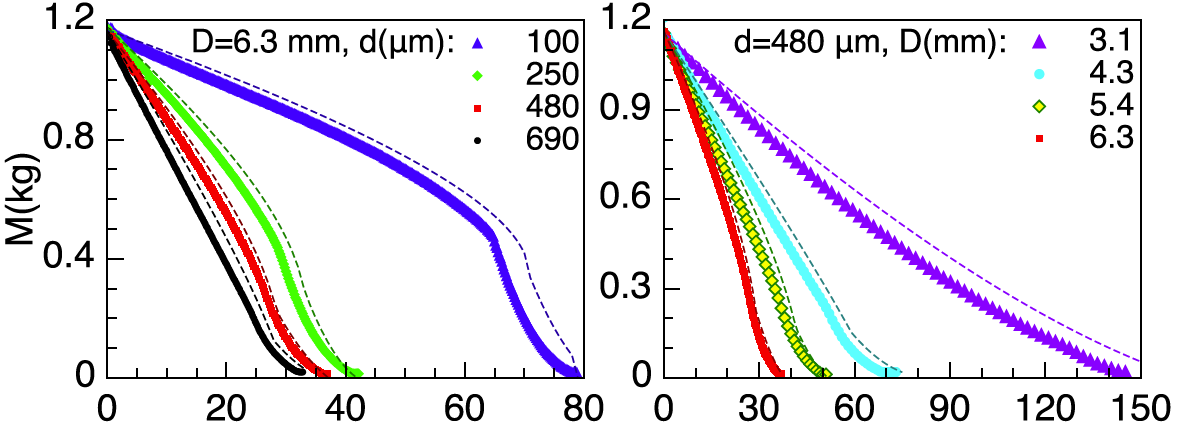}}{0.0cm}{-4.5cm}
\topinset{b)}{\includegraphics[width=9.1cm]{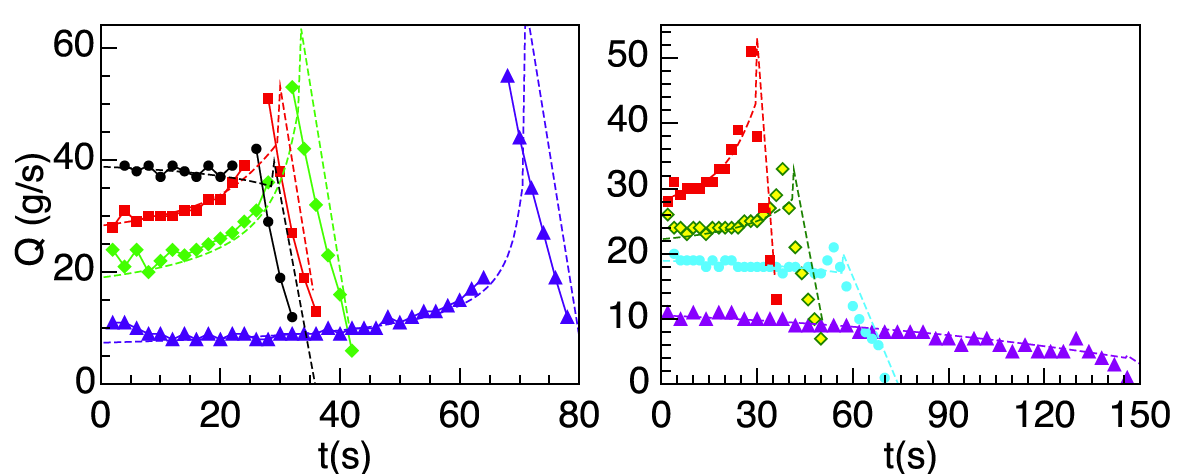}}{0.0cm}{-4.5cm}
\topinset{c)}{\includegraphics[width=9.1cm]{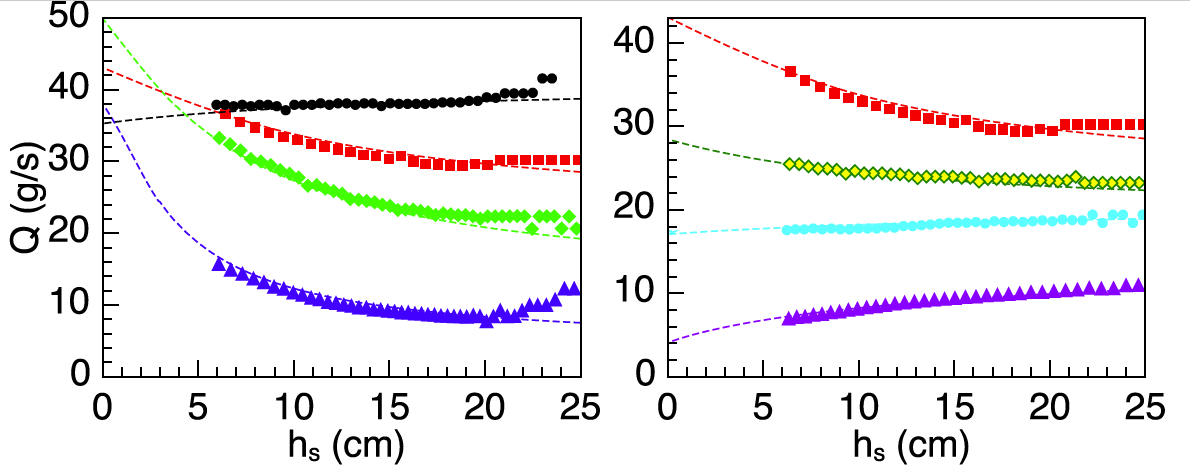}}{0.0cm}{-4.5cm}
\end{center}
\caption{\small (Color online): Comparison of experimental data (points) and numerical results obtained from Eq. \eqref{eq7} (dashed lines) of  a) $M$ vs $t$, b) $Q$ vs $t$, and c) $Q$ vs $h_s$ for different values of  $D$ and $d$. }
\label{fig7}
\end{figure}

Let us now consider the Bernoulli's principle applied to the liquid moving with velocity $v_1=v_l$ at the upper free surface $S_1$ and with velocity $v_2$ at the aperture:
\begin{equation}
\rho_l g h_l - \Delta P = \frac{1}{2} \rho_l \left(v_2^2-v_l^2 \right),
\label{eq5}
\end{equation}
Combining Eqs. \eqref{eq4} and \eqref{eq5} one obtains a second expression relating $h_l$ and $h_s$ :
\begin{equation}
\rho_l g h_l -180 \frac{\eta (1-\epsilon)^2}{d^2 \epsilon^3} h_s (v_l-v_s) = \frac{1}{2} \rho_l \left(v_2^2-v_l^2 \right).
\label{eq6}
\end{equation}
Considering Eq. \eqref{eq3} and the continuity equation applied to the liquid: $S_1 v_l=\epsilon S_2 v_2$, where $\epsilon$ was introduced to take into account only the available porous space at the output, we can solve Eq. \eqref{eq6} for $v_2(h_l)$ and then integrate numerically the continuity equation to obtain $h_l(t)$. Since $h_l$ and $\Delta H$ can be written in terms of $h_s$, the resulting expression for $M$ is:
\begin{equation}
M\approx S_1h_s(t)\left[(\rho_s-\rho_l)\phi+ \rho_l\left(1+b\right)\right]+\rho_l S_1 a,
\label{eq7}
\end{equation}
%
where $h_s(t)$ contains all the information about the system. $M(t)$ is plotted in dashed lines and compared with experiments in Fig. \ref{fig7}a for different values of particles and apertures sizes. Note that the main features of the dynamics are captured by the solution until the silo runs out of grains, where Eq. \eqref{eq7} is matched with Eq. \eqref{eq2}. Finally, from the first derivative of $M(t)$ we obtain $Q(t)$ and the corresponding implicit solution for $Q(h)$, see Figs. \ref{fig7}b,c respectively. In the latter case, the model is able to describe well the experiment in the measured range and predicts $Q(h)$ for $h<5$ cm, which was not accessible experimentally due to set-up limitations. 

\noindent \textit{Discussion}: In the Beverloo's law, the square root dependence on gravity comes from the fact that the dry particles are assumed to start a free fall above the aperture in a region of size proportional to the opening diameter, this is $v_s \propto \sqrt{gD}$. Since Eq. \eqref{eq3} can be written as $v_s\approx (1 - 6d/D)v_l = k v_l$ for $d\ll D$, we can assume that the liquid drags the particles increasing their velocity at the aperture, and therefore $Q$ becomes larger respect to the value expected for dry grains. On the other hand, the grains work as a porous medium that generates hydrodynamic resistance to the liquid and reduces $v_l$ respect to the free liquid case which empties at a larger rate. This helps to understand why $Q_{g} < Q_{m} <Q_{l}$.

Furthermore, from Eq. \eqref{eq3} we have that $v_l > v_s$, consequently, the height of the liquid column over the granular bed decreases during the discharge, i.e. the hydrostatic pressure. For $d\sim100$ $\mu$m, liquid and grains levels decrease almost at the same velocity and the pressure above the granular bed remains nearly constant. Both scenarios coincides with the most marked behaviors of $Q(t)$ shown previously in Fig. \ref{fig3}b. Nevertheless, the observed surge of $Q(t)$ cannot be explained by the simple variation of hydrostatic pressure; we need to consider that $h_s$ is also decreasing, i.e., the hydrodynamic resistance. Therefore, the competition between the decrease of hydrostatic pressure and the hydrodynamic resistance given by the first and second term or Eq. \eqref{eq6} determines the whole dependence of $Q(t)$. In the dry case, an analytical expression for $Q_g$ is possible because the granular column height plays no role. In our experiments, Eq. \eqref{eq6} becomes a non-linear differential equation for $h_l$ after introducing the continuity equation, and only the numerical solution can be reported. 

Finally, let us focus on the big difference between $Q_l$ and $Q_m$ at the end of the discharge. In Refs. \citep{Furnbank2004,Furnbank2007}, it was reported that microparticle-laden liquids can be described as an effective fluid, i.e. as a pure liquid with equivalent viscosity. It has been also found  that dense granular suspensions show a diverging viscosity at increasing particle concentration \cite{Zarraga2000,Bonnoit2012} and that the transition from the effective fluid regime to an interstitial regime is given by the grain diameter and volume fraction \citep{Bonnoit2012}. This last conclusion was obtained by analyzing the detachment of drops of granular suspensions using particles ranging from 20 $\mu$m to 140 $\mu$m at different volume fractions. In our experiments, only the case $d=100$ $\mu$m falls in this range. For this grain size $v_s \sim v_l$, which means that the liquid and grains move as a whole and the mixture is discharged as a dense granular suspension.
This framework helps us to associate the discontinuity in the flow rate observed in Fig. \ref{fig3}b  with an abrupt decrease in the effective viscosity of the mixture. When the grains run out, the viscosity of the granular suspension suddenly decreases to the liquid viscosity, and the flow rate augments abruptly up to the expected value for inviscid water. Note in Fig. \ref{fig3}b that $Q(t)$ always behaves similarly for $d\sim100$ $\mu$m in the studied range of $D$, whereas for $d\sim690$ $\mu$m the change in behavior is remarkable. This could be associated to the transition between two regimes, the first dominated by viscosity and the second one by inertia. 

\noindent \textit{Conclusions}: In a discharge race between a hourglass, a clepsydra and a combined device, the mixture of grains and liquid is faster than dry grains but slower than water. In the mixture, the competition of the hydrostatic pressure above the granular substrate and the hydrodynamic resistance produced by the latter determines the relative amounts of grains and liquid that are discharged, which are, unexpectedly, linearly proportional. In addition, by analyzing the thinning of the filament of material flowing from the silo we obtain some insights about the rheological properties of the mixture, at least qualitatively. A deeper analysis of the transition from a viscous to an inertial regime and the dripping to jetting phenomena depending on the $D/d$ will be considered in a further research. 

\begin{acknowledgments}
The authors acknowledge J. M. Salazar for helpful discussions and suggestions. This work was supported by CONACYT Mexico through the Sectorial Found for Research and Education CB-0242085 and VIEP-BUAP projects. 
\end{acknowledgments}

\bibliographystyle{apsrev}

$^*$ Corresponding author: fpacheco@ifuap.buap.mx

\bibliography{Biblio-BB}

\end{document}